\documentclass[conference]{IEEEtran}
\IEEEoverridecommandlockouts
\usepackage{cite}
\usepackage{amsmath,amssymb,amsfonts}
\usepackage{algorithmic}
\usepackage{graphicx}
\usepackage{textcomp}
\usepackage{xcolor}
\def\BibTeX{{\rm B\kern-.05em{\sc i\kern-.025em b}\kern-.08em
    T\kern-.1667em\lower.7ex\hbox{E}\kern-.125emX}}

\begin{document}

\makeatletter
    \newcommand{\linebreakand}{%
      \end{@IEEEauthorhalign}
      \hfill\mbox{}\par
      \mbox{}\hfill\begin{@IEEEauthorhalign}
    }
\makeatother

\title{Exploring Students’ Insights on Creating Videos to Learn Software Design and Engineering}

\author{

\IEEEauthorblockN{Pedro G. Feij\'oo-Garc\'ia\textsuperscript{\textsection}}
\IEEEauthorblockA{\textit{School of Computing Instruction} \\
\textit{Georgia Institute of Technology}\\
Atlanta, GA, USA \\
pfeijoogarcia@gatech.edu}

\and

\IEEEauthorblockN{John D. Centner\textsuperscript{*}}
\IEEEauthorblockA{\textit{School of Computing Instruction} \\
\textit{Georgia Institute of Technology}\\
Atlanta, GA, USA \\
jcentner3@gatech.edu}

\and

\IEEEauthorblockN{Lucas G. Zangari\textsuperscript{*}}
\IEEEauthorblockA{\textit{School of Computing Instruction} \\
\textit{Georgia Institute of Technology}\\
Atlanta, GA, USA \\
lzangari3@gatech.edu}

\linebreakand
\IEEEauthorblockN{Chase G. Wrenn\textsuperscript{*}}
\IEEEauthorblockA{\textit{Department of CISE} \\
\textit{University of Florida}\\
Gainesville, FL, USA \\
chasewrenn@ufl.edu}
\and
\IEEEauthorblockN{Neha Rani}
\IEEEauthorblockA{\textit{Department of CISE} \\
\textit{University of Florida}\\
Gainesville, FL, USA \\
neharani@ufl.edu}
\and
\IEEEauthorblockN{Juan S. Sánchez-Gómez}
\IEEEauthorblockA{\textit{College of Engineering} \\
\textit{Universidad de los Andes}\\
Bogotá D.C., Colombia \\
js.sanchez14@uniandes.edu.co}
\and
\IEEEauthorblockN{Olufisayo Omojokun}
\IEEEauthorblockA{\textit{School of Computing Instruction} \\
\textit{Georgia Institute of Technology}\\
Atlanta, GA, USA \\
omojokun@cc.gatech.edu}
}

\maketitle
\begingroup\renewcommand\thefootnote{\textsection}
\footnotetext{Corresponding author}
\endgroup
\begingroup\renewcommand\thefootnote{*}
\footnotetext{All authors contributed equally to this research.}
\endgroup

\begin{abstract}
This research-to-practice study examines the use of student-created instructional videos as an active learning strategy in a large undergraduate Software Design and Engineering (SWE) course. Supporting engagement and understanding in large-scale computer science (CS) courses remains a challenge, as instruction often relies on automated activities and summative assessments that prioritize grading over meaningful learning. This can make the learning experience feel impersonal, leading to frustration, anxiety, and confusion, especially when students face complex topics or new programming tools. To address these issues, we implemented a learner-centered strategy grounded in the principle of ``learning by teaching," where students created instructional videos to explain course concepts in their own words. This approach allowed students to take ownership of their learning, adapting the material to fit their individual perspectives and reinforcing their understanding through explanation. We report on students' reflections from a large introductory SWE course (over 700 students), analyzing their open-ended feedback on the benefits and drawbacks of the strategy. Students highlighted increased engagement and learning, and many appreciated the flexibility of the video-based format compared to traditional multiple-choice exams, which they found more restrictive. However, students also noted that the video creation process was time-consuming. This study contributes to the growing conversation on active learning in computing and engineering education by showcasing the potential of student-created instructional videos to enhance student agency, motivation, and conceptual understanding in large-scale learning environments.
\end{abstract}


\section{Introduction}
Active learning in large-scale classrooms presents significant challenges, particularly when instructors simultaneously respond to hundreds of students. As class sizes increase, it becomes increasingly difficult to understand and support students' learning experiences. This makes it hard to follow up on their processes and often forces instruction to rely on automated, standardized assessments that prioritize grading over significant learning \cite{fink2013,magnussen}. Such approaches neglect the importance of formative, learner-centered environments, which require time, intentionality, and personalization to help students learn and, essentially, learn how to learn.

The impersonal nature of large-scale courses can lead to adverse outcomes, especially in computer science (CS), where many students struggle with the abstract nature existing in its curricula \cite{lister, luxton}. Introductory programming courses, in particular, are widely perceived as difficult \cite{robins} and have been shown to contribute to high rates of anxiety and frustration \cite{morgan}. These experiences often lead students to withdraw from the course or even the discipline entirely. The widespread reliance on ``hidden-box" assessments, where students are evaluated based on final outputs rather than their thought processes, further contributes to the development of fragile or inaccurate mental models of programming \cite{gardner}. As a result, many students leave introductory courses with misconceptions \cite{sorva}, which can foster challenges and difficulties in advanced classes for both learners and instructors.

In light of the challenges we have observed when teaching software design and engineering (SWE) content, we developed and evaluated a learner-centered strategy implemented in the Spring of 2024 at a major Southeastern institution in the United States (USA)--the Georgia Institute of Technology. Our strategy is based on the concept of ``learning by teaching," where students create their own instructional videos to demonstrate course concepts. By giving students the opportunity to design and produce these videos, we aimed to empower them to take ownership of their learning by tailoring the format and style to their individual needs and learning preferences.

This research-to-practice paper contributes to discussions on active learning strategies in computing and engineering education, emphasizing the advantages of student-created instructional videos. This is by sharing our efforts to integrate student-created instructional videos into a large-scale undergraduate SWE course with over 700 students, reporting on students' insights and experiences in two scenarios: creating instructional videos or responding to traditional multiple-choice exams. For the scope of this paper, our analysis is centered on qualitative findings. The quantitative results from our strategy will be featured in a separate publication.

\section{Background and Related Work}

Using student-created videos has gained attention as a valuable educational strategy, as they have been seen to foster engagement and learning outcomes in different academic settings, particularly in computer science education (CS Ed). Arzaga et al. \cite{arzaga2023} found that student-created videos foster a sense of ownership and accountability, increasing student engagement and motivation. Students who created their own videos showed significantly better grades than those who used pre-existing YouTube videos, particularly in virtual classroom environments \cite{arzaga2023}.

Allowing students to control the content, format, and style of their videos has been seen to further enhance their learning experiences by aligning them with their preferences, which boosts their motivation and interest \cite{vasilchenko2020}. Student agency on their videos has been found to be particularly beneficial, with students expressing a preference for video creation and a desire for more opportunities to engage in such activities in future instances \cite{gallardo2020}. Additionally, promoting choice and collaboration in video projects has been seen to foster student motivation and the development of conceptual and practical skills \cite{forbes2023}.

Student-created videos have also been found to help students develop a diverse range of skills. Ariza et al., \cite{ariza2023} reported substantial improvements in coding and conceptual assessments due to the creation of video tutorials during students' learning experiences. Feijóo-García and Gardner-McCune \cite{feijoo2020} noted positive impacts on student engagement, learning outcomes, and the overall teaching and learning experience in upper-level computer science courses. Moreover, students have indicated the development of collaboration, communication, information, and media literacy skills as a result of the video creation process \cite{vasilchenko2020}. 

Students' creation of instructional videos has also been shown to enhance retention and build their confidence, particularly in programming skills \cite{vasilchenko2020}. The self-flipped classroom model, which encourages peer teaching through video creation, has been effective in improving communication, project management, and technological skills among students \cite{gallardo2019}. Additionally, peer review is an integral part of the self-flipped classroom model, promoting reflective thinking and collaborative learning by encouraging students to engage critically with the videos created by their peers \cite{gallardo2019}. This reflective process allows students to consider their ideas more deeply and approach their work from new perspectives \cite{campbell2018}.

Although generally beneficial, the strategy of using student-created videos has also been seen to come with some drawbacks. From a faculty perspective, while this approach grants students greater control over how they gain knowledge, it also demands increased effort: A balance that often goes unrecognized by students \cite{daniels2021}. Another concern is the variable quality of student-created videos, which can adversely affect learning outcomes \cite{vasilchenko2018}: Students sometimes focus excessively on the creative aspects of video production at the expense of adequately connecting with the academic content \cite{vasilchenko2018}. Additionally, many students have reported a need for more guidance throughout their learning processes, which can lead to dissatisfaction with the strategy \cite{daniels2021}. The reliance on unverified third-party sources also poses a risk, potentially placing learning outcomes beyond the students' control \cite{daniels2021}.

\section{Educational Strategy} \label{Strategy}
\subsection{Strategy Design}
The educational strategy was implemented during the Spring of 2024 at a large public university in the Southeast of the United States (USA)--the Georgia Institute of Technology. The strategy served as the midterm assessment for three sections of an introductory undergraduate course in Software Design and Engineering (SWE), with participation from over 750 students. The assessment was completed individually and was designed to help students reinforce the concepts covered up to that point in the course. The students were given ten days to complete the task.

The core of the strategy involved students creating instructional videos as the main format of the assessment. Instead of writing answers, students were asked to verbally explain their reasoning when responding to each question, encouraging reflection and articulation of their understanding.

\subsubsection{Question Format}
The assessment consisted of five open-ended questions: two focused on architectural patterns and styles, one on refactoring and code smells, one on requirements engineering, and one on software design principles.

Each question explicitly asked for an explanation, examples, and application. E.g., \textit{``First question: Describe and compare the MVC and MVVM architectural patterns. Use one or more examples to showcase their similarities and differences."}

In addition to the five conceptual questions, we asked a sixth question that had students create a domain model diagram (DMD) based on a provided context description. The students were instructed to explain their reasoning and describe how the elements of the diagram were derived from the context. The diagram had to be hand-drawn, either using a tablet or on paper.

\textit{``Sixth question: Create a Domain Model diagram from the following description and describe how they relate to one another. Explain and elaborate on how you came up with the diagram from the text below. For this question, you must draw the diagram by hand (you can use a tablet or a piece of paper)."}

To promote student engagement, the context scenario was based on the Marvel Universe (highly popular in North America because of comics, movies, and series). Here is how the guideline introduced the context:
\textit{``In the structured yet dynamic Marvel Universe, several unique entities coexist, each defined by specific attributes and playing distinct roles. We need your team to help design the Battle Manager System (BMS), to help manage how battles take place in this fabulous Universe…”}

\subsubsection{Deliverable Requirements} \label{videos}
Students were instructed to create a slide-based presentation (e.g., PowerPoint), responding to each question. They were asked to clearly say the number of the question out loud before beginning their explanation. PowerPoint was recommended due to its built-in voice recording feature, but students were also allowed to use alternatives such as Zoom if preferred.

All submissions were audio-only (no video of the speaker), and students were required to submit a media file not exceeding 500 MB due to constraints in the Learning Management System (LMS) used by the institution. The final video length could not exceed 25 minutes.

\subsection{Strategy Evaluation}
To evaluate the strategy, we conducted a study using an online, asynchronous questionnaire delivered via Qualtrics (see Section \ref{questionnaire}). The goal was to assess how engaging and useful this formative approach could be in comparison to more traditional assessment methods commonly used in college.

We designed three slightly different versions of the strategy across the three course sections. The study was approved by the Institutional Review Board (IRB) at the Georgia Institute of Technology (Protocol No H24031). Participation was voluntary, and students who chose to take part were given extra-credit as an incentive.

Student-created instructional videos were reviewed by the instructional team of the course, consisting of a leading instructor (corresponding author) and 24 undergraduate teaching assistants (UTAs). This was a variation from our previous pilot from the Fall of 2023 (from August to December), which considered students peer-reviewing the instructional videos \cite{iticse_videos}. For this time, grades were not considered and were not analyzed as our focus concerned students' experiences and engagement toward our learning strategy and to protect student anonymity for ethical purposes.

\subsubsection{Method} \label{method}
We did not announce how the midterm exam was going to be (i.e., using student-created videos). However, a week before the midterm week, we encouraged our students to review the course content and create a cheat sheet they could bring to the proposed assessment. The assessment was described as an in-class close-ended multiple-choice in-class test via their Learning Management System (LMS). 

Here are the approaches we compared:

\begin{enumerate}
    \item First Section: Test With Scores + Videos
    \begin{enumerate}
        \item First, students responded to an in-class multiple-choice close-ended test of 20 minutes and 20 questions. We configured the test so students could see their scores but not the right answers. All questions were True/False and concept-focused.
        \item They were then asked to respond to the online questionnaire (see Section \ref{questionnaire})  with their perceptions and thoughts.
        \item Students were then asked to work on our educational strategy and create their videos (see Section \ref{videos}).
        \item After the deadline and before grades came by, students responded to the online questionnaire (see Section \ref{questionnaire})  with their perceptions and thoughts.
    \end{enumerate}
\item Second Section: Test With no Scores + Videos
    \begin{enumerate}
        \item First, students responded to an in-class multiple-choice close-ended test of 20 minutes and 20 questions. We configured the test so students could not see their scores or the right answers. All questions were True/False and concept-focused.
        \item They were then asked to respond to the online questionnaire (see Section \ref{questionnaire})  with their perceptions and thoughts.
        \item Students were then asked to work on our educational strategy and create their videos (see Section \ref{Strategy}).
        \item After the deadline and before grades came by, students responded to the online questionnaire (see Section \ref{questionnaire})  with their perceptions and thoughts.
    \end{enumerate}
\item Third Section: Videos (Exclusively)
    \begin{enumerate}
        \item Students were only asked to work on our educational strategy and create their videos (see Section \ref{Strategy}).
        \item After the deadline and before grades came by, students responded to the online questionnaire (see Section \ref{questionnaire})  with their perceptions and thoughts.
    \end{enumerate}
\end{enumerate}

\subsubsection{Questionnaire} \label{questionnaire}
The questionnaire had three parts in total:

\begin{enumerate}
\item \textbf{The first part} presented an informed consent document with three questions:
    \begin{enumerate}
        \item Confirmation of majority of age.
        \item Voluntary participation agreement.
        \item An open-ended asking to input the student's institutional email address for communication and extra-credit purposes.
    \end{enumerate}
    \item \textbf{The second part} featured five questions:
    \begin{enumerate}
        \item A close-ended question on a 7-point semantic scale about the student's perceptions on their performance.
        \item A close-ended question with six 7-point Likert scales regarding the student's perceptions of the assessment strategy and their learning process. These question adapted the Liker scales from Diemer et al., measuring students' \textbf{\textit{perceptions of learning}} \cite{diemer2012student}.
        \item An open-ended question about positive aspects of the assessment strategy: ``Tell us what you \textbf{\textit{liked}}..."
        \item An open-ended question about negative aspects of the assessment strategy: ``Tell us what you \textbf{\textit{disliked}}..."
        \item An open-ended question asking students to estimate hours spent preparing for and completing the assessment strategy.
    \end{enumerate} 
    \item \textbf{The third part} featured 11 questions about students' age, ethnicity, gender, academic major, student origin (domestic to the United States or international), academic level (undergraduate or graduate), academic year, number of credits for the Spring of 2024, number of extracurricular activities, English proficiency, and frequently used languages.
\end{enumerate}

\subsubsection{Data Analysis} \label{analysis}
For the scope of this paper, our analysis is \textbf{qualitative} and centers on the open-ended questions proposed in the questionnaire to gather positive and negative insights about the type of assessment used (see Section \ref{questionnaire}). We analyzed responses to open-ended questions and participants' voices with a thematic analysis \cite{thematic}--method used for the identification and interpretation of patterns and the affinity of participants' experiences and insights.

We organized our analysis considering the following:
\begin{enumerate}
    \item Positive input in the in-class multiple choice test (first and second sections).
    \item Negative input on the multiple choice in-class test (first and second sections).
    \item Positive input on the video-creation educational strategy (All three sections).
    \item Negative input on the video-creation educational strategy (All three sections).
\end{enumerate}

The responses to our closed questions will be analyzed and reported in a separate manuscript.

\subsection{Student Population}
We had a total of N=787 students across the three sections we report on (see Section \ref{method}). Out of those, n=657 students (83.5\%) voluntarily decided to share their thoughts and perceptions with us by answering our questionnaire.

Most of our students were in the early stages of their academic programs. A total of 86.3\% were in their first (n=218) or second (n=349) academic year. In terms of course load, 72.1\% of the students reported taking between 12 and 16 academic credits, although most students were enrolled in between 12 to 19 credits.

The majority of students were between 18–20 years old (n=611), followed by students in the 21–25 age range (n=45), and one student (n=1) who reported being 30 years old.

Almost all students who responded (n=641--97.6\%) were pursuing computing-related majors, such as computer science, computer engineering, or computational media. The course is mandatory for students in those majors and for students pursuing a minor in computing.

In terms of gender identity, we had males (n=480), females (n=163), and students who identified in other gender groups (n=14). Regarding ethnicity, students self-identified as Asian (n=421), Afro/Black American (n=37), Hispanic/Latin American (n=24), White American—non-Hispanic/non-Latin American (n=121), and Middle Eastern/North African (n=15). A total of n=39 students reported belonging to two or more ethnic groups.

We had both domestic (n=566) and international (n=91) students. A majority of students were multilingual (n=523), while others identified as monolingual (n=135). In terms of language proficiency, students reported being either native English speakers (n=437) or English learners (n=220).

\section{Findings}
This section presents our outcomes based on the data analysis strategy suggested in Section \ref{analysis}.

\subsection{Insights on Students' Experiences: Multiple-Choice Question Modality}

Students were asked to share their insights from their experiences with the multiple-choice question modality. We report on responses from the First and Second sections (n=494). Table \ref{table:testPositive} displays the nine themes identified through the thematic analysis conducted on students' insights on what they liked about the modality.

\begin{table}[ht]
\centering
\caption{\centering\label{table:testPositive}Themes From What Students Liked About the Multiple-Choice Format}
\begin{tabular}{|c|l|l|}
\hline
\textit{\textbf{ID}} & \multicolumn{1}{c|}{\textit{\textbf{Theme}}} & \multicolumn{1}{c|}{\textit{\textbf{Description}}} \\ \hline
\textit{TL\_1} & \textit{\begin{tabular}[c]{@{}l@{}}Preparation \\ and Learning\end{tabular}} & \textit{\begin{tabular}[c]{@{}l@{}}Used for responses in which \\ participants indicated they learned, \\ they could prepare for \\ the test, they had a positive \\ experience with the sheet cheat, \\ or responses suggesting \\ a positive learning experience.\end{tabular}} \\ \hline
\textit{TL\_2} & \textit{Question Format} & \textit{\begin{tabular}[c]{@{}l@{}}Used for responses in which \\ participants positively perceived the \\question format.\end{tabular}} \\ \hline
\textit{TL\_3} & \textit{Assessment Topics} & \textit{\begin{tabular}[c]{@{}l@{}}Used for responses in \\ which participants \\ referred to the question's \\ content (or what was asked).\end{tabular}} \\ \hline
\textit{TL\_4} & \textit{Navigation Format} & \textit{\begin{tabular}[c]{@{}l@{}}Used for responses that \\ indicated positive aspects \\ of having a test sequence \\ that did not allow students \\ to return to previous questions.\end{tabular}} \\ \hline
\textit{TL\_5} & \textit{Time Management} & \textit{\begin{tabular}[c]{@{}l@{}}Used for responses that \\ referred to time.\end{tabular}} \\ \hline
\textit{TL\_6} & \textit{Question Clarity} & \textit{\begin{tabular}[c]{@{}l@{}}Used for responses that \\ indicated anything related \\ to the wording or clarity \\ of a question.\end{tabular}} \\ \hline
\textit{TL\_7} & \textit{Positive Experience} & \textit{\begin{tabular}[c]{@{}l@{}}Used for responses in \\ which participants explicitly \\ expressed a positive experience.\end{tabular}} \\ \hline
\textit{TL\_8} & \textit{Negative Experience} & \textit{\begin{tabular}[c]{@{}l@{}}Used for responses in \\ which participants explicitly \\ expressed a negative experience.\end{tabular}} \\ \hline
\textit{TL\_9} & \textit{Empty} & \textit{\begin{tabular}[c]{@{}l@{}}Used when none of \\ the other categories \\ matched a response.\end{tabular}} \\ \hline
\end{tabular}
\end{table}

Students appreciated the possibility of using a cheat sheet for the multiple-choice modality, making \textit{\textbf{Preparation and Learning}} as the theme with the highest frequency (43.1\% of responses indicated it--n=213). In general, students indicated the cheat sheet helped them prepare for the assessment, as they found it helped them foster their understanding of the topics: 

\begin{itemize}
    \item \textit{``Doing the cheat-sheet definitely helped me better remember the content, even when I wasn't checking it, so that was nice."} [B059].
    \item \textit{``The allowance to bring cheat sheet made me actually understand what I was writing and learn the material"} [B062].
    \item \textit{``I liked that we were able to bring a sheet of paper to aid us with the mini-assessment. Because our paper had to be handwritten, creating the ``cheat sheet" was actually a method of studying by actively reading/writing down information." }[C236].
    \item \textit{``I liked the use of being able to bring a study sheet to the exam, because it helped me focus more on understanding concepts than memorizing content."} [C018].
\end{itemize}

Following, \textit{\textbf{Question Format}} made it the second most popular theme (30.8\% of responses--n=152), with students particularly appreciating the True/False questions used during the modality. In general, students indicated the question format helped to prevent ambiguity, as well as test their understanding of topics: 

\begin{itemize}
    \item \textit{``True or false questions forced me to ensure that the statement wasn't just kind of correct, but fully correct. This really tested the level of depth of my understanding regarding the content covered in lectures."} [B003].
    \item \textit{``I liked that it was only a T/F question, leaving less room for ambiguity."} [B084].
    \item \textit{``Everything being true or false made me think about edge cases or formalities in the definitions and applications of concepts, which really made me think hard about what I've learned." }[C178].
    \item \textit{``I liked that all the questions were true or false, since it better consolidated certain concepts. I don't think the questions were too nuanced, so it worked fine. The questions broadly covered all the content we learned so far, so that was good too."} [C234].
\end{itemize}

\begin{table}[ht]
\centering
\caption{\centering\label{table:testNegative}Themes From What Students Disliked About the Multiple-Choice Format}
\begin{tabular}{|c|l|l|}
\hline
\textit{\textbf{ID}} & \multicolumn{1}{c|}{\textit{\textbf{Theme}}} & \multicolumn{1}{c|}{\textit{\textbf{Description}}} \\ \hline
\textit{TD\_1} & \textit{\begin{tabular}[c]{@{}l@{}}Time and Effort\end{tabular}} & \textit{\begin{tabular}[c]{@{}l@{}}Used for responses commenting \\on a negative experience regarding \\time and their commitment \\to the assessment.\end{tabular}} \\ \hline
\textit{TD\_2} & \textit{Navigation Format} & \textit{\begin{tabular}[c]{@{}l@{}}Used for negative experiences \\ concerning the sequential format \\ of the assessment.\end{tabular}} \\ \hline
\textit{TD\_3} & \textit{Assessment Content} & \textit{\begin{tabular}[c]{@{}l@{}}Used for negative experiences \\ concerning the topics asked \\ in the assessment.\end{tabular}} \\ \hline
\textit{TD\_4} & \textit{Question Format} & \textit{\begin{tabular}[c]{@{}l@{}}Used for negative experiences with\\ the question format: True/False \\Questions.\end{tabular}} \\ \hline
\textit{TD\_5} & \textit{Question Clarity} & \textit{\begin{tabular}[c]{@{}l@{}}Used for negative experiences \\ concerning the clarity or wording \\ of the questions.\end{tabular}} \\ \hline
\textit{TD\_6} & \textit{Lack of Engagement} & \textit{\begin{tabular}[c]{@{}l@{}}Used for responses indicating that \\ the participant did not give their \\ best or 100\% to the assessment.\end{tabular}} \\ \hline
\textit{TD\_7} & \textit{Uncertainty} & \textit{\begin{tabular}[c]{@{}l@{}}Used when students indicate \\ they feel uncertain or overthought \\ their responses.\end{tabular}} \\ \hline
\textit{TD\_8} & \textit{Positive Experience} & \textit{\begin{tabular}[c]{@{}l@{}}Used for responses in which \\participants suggest a positive \\experience, referring positively \\to stress, difficulty, engagement, \\ or overall satisfaction.\end{tabular}} \\ \hline
\textit{TD\_9} & \textit{Negative Experience} & \textit{\begin{tabular}[c]{@{}l@{}}Used for responses in which \\participants suggest a negative \\experience.\end{tabular}} \\ \hline
\textit{TD\_10} & \textit{Empty} & \textit{\begin{tabular}[c]{@{}l@{}}Used when none of the other \\ categories match the response.\end{tabular}} \\ \hline
\end{tabular}
\end{table}

Table \ref{table:testNegative} displays the ten themes identified through the thematic analysis conducted on students' insights on what they disliked about the multiple-choice modality. The category mostly present among students' responses (n=454) was \textit{\textbf{Navigation Format}} (33.0\% of responses--n=163), generally criticizing the incapacity students had during the assessment to return to previous questions after responding to them:

\begin{itemize}
    \item \textit{``I couldn't go back on my questions and look through my answers one last time for if I made mistakes."} [B040].
    \item \textit{``I did not like the locked questions, as I typically go back to check my work."} [B157].
    \item \textit{``I did not like how I was unable to go back to a question after I moved past it. Oftentimes when I take exams/quizzes I will remember a key piece of information later in the quiz that causes me to alter an answer to a previous question. With this style, I was unable to do that. Additionally, I was unable to skip a question I was confused on and come back to it after answering ones that I felt confident in." }[C239].
    \item \textit{``I didn't like that you couldn't go back and forth between questions, as it made it difficult to keep track of time and also not spend too long on previous questions before proceeding."} [C253].
\end{itemize}

\textit{\textbf{Assessment Content}} made it the top two themes (30.6\% of responses--n=151), with students particularly disliking how the assessment's content was more theoretical than practical, with questions perceived too specific:

\begin{itemize}
    \item \textit{``The assessment seemed to be more factual, than problem-solving based. Although, the course content certainly steers towards that direction."} [B125].
    \item \textit{``It was very hard to remember some stuff that were tested, because the questions were too specific in the nature they were asked."} [B141].
    \item \textit{``I don't think a quiz is necessarily the best way to assess our understanding of the ideas, because it's mostly just memorization and not based on understanding." }[C027].
    \item \textit{``I disliked how the assessment tested more memorization of content than anything."} [C163].
\end{itemize}

\textit{\textbf{Question Clarity}} also made it to the second place (30.6\% of responses--n=151), with students criticizing the wording and specificity of questions. Some students even referred to language barriers based on English not being their main language:

\begin{itemize}
    \item \textit{``Some questions are too wordy like it I wish rather than just focusing on tiny details we could have questions more on applications"} [B030].
    \item \textit{``It is not easy for me to understand the question itself as English is not my mother language, hope there will be clear and simple words"} [B192].
    \item \textit{``I disliked that some of the questions were slightly vague in their wording." }[C019].
    \item \textit{``I did not like how sometimes it was worded weird as it made the problem itself difficult to understand."} [C088].
\end{itemize}

Overall, students welcomed and regretted aspects that affected their learning agency during the multiple-choice modality. The cheat sheet and question format reflected on students' interest towards having control of their learning processes, appreciating a method that reduced uncertainty or as [B084] indicated, ``\textit{ambiguity}." On the other hand, the sequential navigation format was generally disliked among students, as it was seen as a limitation that removed their control towards their responding process. 

Likewise, as students commented, aspects of the assessment were criticized for clarity and content. From what we observed, students would have appreciated questions that prevented misunderstandings (i.e., ambiguity) and additional means to elaborate on their responses and integrate their knowledge beyond the test.

\subsection{Insights on Students' Experiences: Video Creation Modality}
Students were also asked about their experiences with the video creation modality. We report on responses from all three sections (n=540). Table \ref{table:videoPositive} displays the 12 themes identified through the thematic analysis conducted on students' insights on what they liked about the modality.

\begin{table}[ht]
\centering
\caption{\centering\label{table:videoPositive}Themes From What Students Liked About Creating Videos}
\begin{tabular}{|c|l|l|}
\hline
\textit{\textbf{ID}} & \multicolumn{1}{c|}{\textit{\textbf{Theme}}} & \multicolumn{1}{c|}{\textit{\textbf{Description}}} \\ \hline
\textit{VL\_1} & \textit{Assessment Flexibility} & \textit{\begin{tabular}[c]{@{}l@{}}Used for responses that refer \\to the format of the assessment \\(e.g., take-home or open-ended).\end{tabular}} \\ \hline
\textit{VL\_2} & \textit{Time Management} & \textit{\begin{tabular}[c]{@{}l@{}}Used for responses that referred \\to time.\end{tabular}} \\ \hline
\textit{VL\_3} & \textit{Preparation and Study} & \textit{\begin{tabular}[c]{@{}l@{}}Used for responses that referred 
\\to how the assessment (during \\the process of completing it)\\ helped students explore topics, \\ prepare better, read more, etc.\end{tabular}} \\ \hline
\textit{VL\_4} & \textit{\begin{tabular}[c]{@{}l@{}}Understanding and\\ Learning\end{tabular}} & \textit{\begin{tabular}[c]{@{}l@{}}Used for responses in which \\students explicitly referred to how\\ they learned more or gained \\a better understanding.\end{tabular}} \\ \hline
\textit{VL\_5} & \textit{Response Flexibility} & \textit{\begin{tabular}[c]{@{}l@{}}Used for responses in which\\ students referred to having \\agency or the capacity \\to adapt their responses \\with their own examples or ideas.\end{tabular}} \\ \hline
\textit{VL\_6} & \textit{Learned by Teaching} & \textit{\begin{tabular}[c]{@{}l@{}}Used for responses in which\\ students explicitly indicated that\\ explaining or teaching helped \\them learn more.\end{tabular}} \\ \hline
\textit{VL\_7} & \textit{Assessment Interactivity} & \textit{\begin{tabular}[c]{@{}l@{}}Used for responses that\\ indicated that the video\\ format of the assessment was \\more interactive, positive, or\\ better than other formats.\end{tabular}} \\ \hline
\textit{VL\_8} & \textit{Assessment Content} & \textit{\begin{tabular}[c]{@{}l@{}}Used for responses that\\ referred to the type of \\ questions asked during the\\ assessment.\end{tabular}} \\ \hline
\multicolumn{1}{|l|}{\textit{VL\_9}} & \textit{Assessment Deliverable} & \textit{\begin{tabular}[c]{@{}l@{}}Used for responses that\\ referred to the deliverable of the \\assessment and its conciseness.\end{tabular}} \\ \hline
\multicolumn{1}{|l|}{\textit{VL\_10}} & \textit{Positive Experience} & \textit{\begin{tabular}[c]{@{}l@{}}Used for responses in which\\ participants explicitly expressed a \\positive experience.\end{tabular}} \\ \hline
\multicolumn{1}{|l|}{\textit{VL\_11}} & \textit{Negative Experience} & \textit{\begin{tabular}[c]{@{}l@{}}Used for responses in which\\ participants explicitly expressed a \\negative experience.\end{tabular}} \\ \hline
\textit{VL\_12} & \textit{Empty} & \textit{\begin{tabular}[c]{@{}l@{}}Used when none of the other \\categories matched a response.\end{tabular}} \\ \hline
\end{tabular}
\end{table}

\begin{table}[t]
\centering
\caption{\centering\label{table:videoNegative}Themes From What Students Disliked About Creating Videos}
\begin{tabular}{|c|l|l|}
\hline
\textit{\textbf{ID}} & \multicolumn{1}{c|}{\textit{\textbf{Theme}}} & \multicolumn{1}{c|}{\textit{\textbf{Description}}} \\ \hline
\textit{VD\_1} & \textit{Time and Effort} & \textit{\begin{tabular}[c]{@{}l@{}}Used for responses that described\\ negative experiences related to the \\ time commitment or effort required \\to complete the assessment.\end{tabular}} \\ \hline
\textit{VD\_2} & \textit{Response Format} & \textit{\begin{tabular}[c]{@{}l@{}}Used for responses criticizing\\ the format or method required for \\presenting or recording responses.\end{tabular}} \\ \hline
\textit{VD\_3} & \textit{Ineffective Learning} & \textit{\begin{tabular}[c]{@{}l@{}}Used for responses suggesting\\ the assessment was not \\a useful or effective learning \\tool.\end{tabular}} \\ \hline
\textit{VD\_4} & \textit{Grading Uncertainty} & \textit{\begin{tabular}[c]{@{}l@{}}Used for responses expressing\\ concerns or confusion regarding\\ the rubric, criteria, or\\ overall grading process \\of the assessment.\end{tabular}} \\ \hline
\textit{VD\_5} & \textit{Assessment Content} & \textit{\begin{tabular}[c]{@{}l@{}}Used for responses that negatively\\ commented on the topics\\ covered in the assessment, \\indicating that the content was \\irrelevant, inappropriate, or \\uninteresting.\end{tabular}} \\ \hline
\textit{VD\_6} & \textit{Question Format} & \textit{\begin{tabular}[c]{@{}l@{}}Used for responses criticizing \\the format of the question \\(e.g., asking for examples, \\ being open-ended, etc.).\end{tabular}} \\ \hline
\textit{VD\_7} & \textit{Question Clarity} & \textit{\begin{tabular}[c]{@{}l@{}}Used for responses highlighting \\issues with the wording or \\clarity of the questions.\end{tabular}} \\ \hline
\textit{VD\_8} & \textit{Exploration} & \textit{\begin{tabular}[c]{@{}l@{}}Used for responses describing \\negative experiences related to the \\need for further exploration\\or research to answer the \\assessment questions.\end{tabular}} \\ \hline
\multicolumn{1}{|l|}{\textit{VD\_9}} & \textit{\begin{tabular}[c]{@{}l@{}}Assessment \\ Restrictions\end{tabular}} & \textit{\begin{tabular}[c]{@{}l@{}}Used for responses describing\\ negative experiences caused by\\ specific restrictions imposed \\by the assessment.\end{tabular}} \\ \hline
\multicolumn{1}{|l|}{\textit{VD\_10}} & \textit{Technical Issues} & \textit{\begin{tabular}[c]{@{}l@{}}Used for responses describing\\ technical problems encountered \\during the assessment, such as\\ software glitches, hardware \\malfunctions, or internet \\connectivity issues.\end{tabular}} \\ \hline
\multicolumn{1}{|l|}{\textit{VD\_11}} & \textit{\begin{tabular}[c]{@{}l@{}}Pre-assessment\\ Instructions\end{tabular}} & \textit{\begin{tabular}[c]{@{}l@{}}Used for responses criticizing\\ the instructions or guidance\\ provided before the \\assessment,  indicating that they\\ were unclear, insufficient, \\or misleading.\end{tabular}} \\ \hline
\textit{VD\_12} & \textit{Positive Experience} & \textit{\begin{tabular}[c]{@{}l@{}}Used for responses in which\\ participants explicitly expressed\\ a positive experience.\end{tabular}} \\ \hline
\multicolumn{1}{|l|}{\textit{VD\_13}} & \textit{Negative Experience} & \textit{\begin{tabular}[c]{@{}l@{}}Used for responses in which\\ participants explicitly expressed a \\negative experience.\end{tabular}} \\ \hline
\multicolumn{1}{|l|}{\textit{VD\_14}} & \textit{Empty} & \textit{\begin{tabular}[c]{@{}l@{}}Used when none of the other \\categories matched a response.\end{tabular}} \\ \hline
\end{tabular}
\end{table}

Students highlighted the benefits that the creation of the instructional videos had on their learning processes. \textit{\textbf{Understanding and Learning}} was the theme with the highest frequency of appearances (59.1\% of responses indicated it--n=319), with students appreciating the oral format of the modality, how it let them elaborate on examples, and the agency they had on the information they needed to used during the assessment:

\begin{itemize}
    \item \textit{``I liked the format of the assessment. Its format is similar to teaching someone else the material. In order to teach someone the material, I would need to first understand it."} [B088].
    \item \textit{``I felt I retained a lot more of what I learned in phase 2 than in phase 1. Phase one felt like mindless memorization and the material soon slipped away. In phase 2 I spent more time understanding what I was going to present about and now feel like I have a better grasp of the content."} [C122].
    \item \textit{``I enjoyed how we had to record because saying it out loud helps register information in my head."} [D029].
\end{itemize}

Particularly, concerning student agency, \textbf{\textit{Response Flexibility}} (24.1\% of responses--n=130) and \textbf{\textit{Assessment Flexibility}} (20.4\% of responses--n=110) were noticeable themes. Students appreciated having freedom to manage the time and structure of their explanations, perceiving the modality as low on stress and engaging:

\begin{itemize}
    \item \textit{``I really learned a lot compared to phase one. I think this is because I used examples to gain a higher level understanding, rather than focusing on small details for each architecture like I did for part one."} [C115].
    \item \textit{``I liked that I could explain the content in my way how I understood the content."} [C247].
    \item \textit{``I liked that it let the student have control over how they wanted to explain the answers to questions in their own words."} [D075].
    \item \textit{``I liked that it forced us to demonstrate that we actually learned the material instead of memorizing it. Also it was nice that there was room for creativity when creating the slideshow."} [D135].
    \item \textit{``I liked that we were free to design the presentation however we wanted. The instructions were not restrictive, which allowed me to design and answer the questions how I wanted to."} [B117].
    \item \textit{``I liked how I had time to talk about my understanding of my own work in my own word."} [B144].
\end{itemize}

Moreover, students' responses suggested they were able to learn by teaching. \textit{\textbf{Learned by Teaching}} (18.7\% of responses--n=101) made it to the fourth position among the 12 themes identified. Students referred to the benefits of teaching and how they perceived their understanding was fostered by having to explain and elaborate on their responses for this modality:

\begin{itemize}
    \item \textit{``I liked that it was much more flexible, yet applicable in the content. It's really effective learning to be teaching/explaining the content."} [B128].
    \item \textit{``I think an assessment that involves teaching other people reinforces what you know and what you thought you know but you don't know well enough."} [B158].
    \item \textit{``I overall liked the aspect in which we were able to ``teach" the concepts to someone else, which can overall help our understanding of the concepts learned in class."} [C221].
    \item \textit{``I liked that I got to feel like Dr. Pedro. I was explaining things and speaking over a powerpoint [PowerPoint] just like him, which was fun!"} [C191].
    \item \textit{``Instead of doing a quiz which makes us memorize contents, the mini-assessment reinforced the concepts by forcing us to explain the topics. Teaching others is the highest level of learning, and it helped me store the concepts more SOLIDly [solidly]."} [D020].
    \item \textit{``I liked that it was self learning and reteaching is the best way I learn. I find the Feynman technique the best way to study and this is similar to that. I also liked the final question because it helped with application."} [D076].
\end{itemize}

Table \ref{table:videoNegative} displays the fourteen themes identified through the thematic analysis conducted on students' insights on what they disliked about the video-creation modality. The category most frequently observed among students' responses (n=540) was \textit{\textbf{Time and Effort}} (32.2\% of responses--n=174), with students indicating that the creation of instructional videos implied much work and dedication:

\begin{itemize}
    \item \textit{``I think overall this strategy was good and less stressful, however I also felt that took a decent amount of time to complete thoroughly."} [B172].
    \item \textit{``I disliked having how long the process of getting ready to take the video took (researching information and setting up the ppt [PowerPoint] took a while)"} [B182].
    \item \textit{``I disliked that the assessment was really long because there was so many parts to each question and I had to make sure every art was correctly addressed in the presentation."} [C049].
    \item \textit{``I did not enjoy the fact that the questions required an extensive amount of detail. It meant that there was a lot of time and effort put into the presentation and the script."} [D126].
\end{itemize}

\textit{\textbf{Response Format}} was the second most frequent theme (29.6\% of responses--n=160), followed by \textit{\textbf{Ineffective Learning}} (27.0\% of responses--n=146). 

\begin{itemize}
    \item \textit{``I didn't like how some of the questions asked could be directly answered by reiterating the slides. Instead, I would appreciate it more if the questions required more critical thinking and applying our knowledge of concepts through examples."} [D041].
    \item \textit{``I didn't like that we had to record ourselves in a presentation style.  I also didn't like that we had to write about the same information that we had already learned for the first mini-assessment."} [C201].
    \item \textit{``Completing the first 5 questions (everything but the domain model) was really boring, maybe making it a group thing would help with that."} [C171].
    \item \textit{``I didn’t like the fact that the assignment had to be recorded. I felt that if the assignment was just turning in a powerpoint [PowerPoint], it would have measured each student’s understanding of the subject matter to the same degree."} [C038].
    \item\textit{``I felt like the video aspect of it, rather than the presentation felt a little like busy work and secondary to actually learning the content which the presentation itself helped with more."} [B115].
    \item\textit{``I didn’t like how this was repetitive to Phase 1 [multiple-choice format] and thought it was somewhat unnecessary."} [B108].
\end{itemize}

We observed that a downside was the amount of time and dedication students invested in the modality. Although it provided them with agency, some students found the format time-consuming. Future experiences should consider adapting the length of the assignment with a more flexible time window to ensure students engage without perceiving this modality as time-consuming. Moreover, as some students reported, having less theoretical and more hands-on questions could help with student engagement.

Nonetheless, students appreciated the flexibility that came with creating their instructional videos. The opportunity to elaborate on their responses and adapt them based on their understanding was highlighted by many of them. Moreover, most students (59.1\%--n=319) referred to the video creation modality as helpful in fostering their understanding and learning. As observed, using student-created instructional videos can foster student engagement, boosting their confidence in their learning processes: Students find themselves excited with strategies like this.

\section{Discussion}
This research-to-practice study provides valuable insights into the benefits and challenges of using student-created instructional videos in CS education, informed by students' engagement and perceptions. Our findings suggest that while students recognized limitations in both assessment modalities, they also valued several aspects of each format that supported their learning processes and agency.

On the positive end, students highlighted multiple benefits of the video creation modality. Many appreciated the oral format, the ability to elaborate with examples, and the freedom to choose what content to include in their responses. A large portion of the course viewed the format as engaging and less stressful than traditional exams, as it allowed them to manage their own time and the structure of their explanations. Interestingly, a significant number of students (18.7\%--n=101) explicitly mentioned the learning benefits associated with teaching, reporting that having to explain and elaborate helped reinforce their understanding. This supports the core idea behind our ``learning by teaching" approach and resonates with previous literature on student-created artifacts \cite{vasilchenko2020, feijoo2020}.

When reflecting on the traditional multiple-choice modality, students also shared positive feedback, particularly related to preparation. The most frequently mentioned theme (43.1\%--n=213) was the use of a cheat sheet, which students viewed as a helpful tool to study and strengthen their understanding. The second most appreciated aspect was the question format (30.8\%--n=152), with students highlighting the clarity and precision of the True/False structure. Many noted that this format reduced ambiguity and made it easier to test their knowledge. 

Nonetheless, several concerns were raised about the multiple-choice modality. The most common theme among responses (33.0\%--n=163) was the sequential navigation format, which prevented students from returning to previous questions. This limitation was broadly criticized for removing control over the assessment experience. Additionally, concerns regarding the assessment content (30.6\%--n=151) and question clarity (30.6\%--n=151) were prevalent, with students pointing out that the content felt overly theoretical and some questions were too specific or difficult to interpret. In several responses, students connected this to language barriers and indicated possible limitations to those who were non-native English speakers. These concerns suggest a strong desire for transparency and flexibility.

Although the instructional video modality was positively received, it was not without some drawbacks. The most frequent concern raised was the time and effort required (32.2\%--n=174). Many students found the format demanding, even if they appreciated the flexibility it offered. Response format (29.6\%--n=160) and concerns about its effectiveness as a learning strategy (27.0\%--n=146) also appeared across responses. Some students suggested that reducing the length of the assignment or offering a more flexible timeline could help improve their experience and reduce perceived workload. Others proposed including more hands-on (e.g., coding-based) questions to increase engagement and practical application.

Despite these limitations, students consistently mentioned that creating their own instructional videos helped them better understand the material. In fact, the majority (59.1\%--n=319) identified this modality as beneficial for their learning. These observations reinforce the idea that giving students space to reflect, articulate, and share their understanding (especially through creative strategies) can increase both engagement and confidence.

Overall, our findings highlight the potential of student-created instructional videos to foster meaningful learning experiences in CS education. However, careful consideration must be given to students' situational factors. Future adaptations of this strategy should consider students' available time and other factors, such as their native languages. Instructional video creation can be a powerful tool to support student agency and reflection, but it must be thoughtfully designed to meet the diverse needs we may find in our CS classrooms.

\section{Conclusions}
In this research-to-practice paper, we have shared students' experiences with the use of student-created instructional videos for an undergraduate course in software design and engineering. By using this strategy in our classrooms, we aimed to promote an active learning, learner-centered experience that fostered students' engagement and understanding of course content. Overall, students generally appreciated our strategy due to its flexibility, the agency it offered, and the perceived learning and performance it provided. Moreover, we observed significant findings indicating the positive outcomes of strategies like ours to motivate women students.

Our work contributes to discussions on active learning strategies in computing and engineering education, emphasizing the advantages of student-created instructional videos. Our findings suggest the possibility of integrating student-created artifacts in computing education, with insights on their benefits in students' learning experiences and engagement. 

Future steps may explore how creating student-generated artifacts impacts students' self-efficacy and performance, not only in software design and engineering but also in other educational contexts. These could include introductory programming courses, hardware-based curricula, and broader STEM (i.e., Science, Technology, Engineering, and Mathematics) areas across engineering and sciences. Additionally, future work could consider expanding the types of deliverables students create. With current trends in artificial intelligence (AI) and multi-modal AI agents, future efforts may investigate the educational value of placing students in the role of designers of these technologies and how this affects their learning experiences.

We hope our work serves to promote research into the use of student-created artifacts in STEM education, encouraging educators to incorporate strategies like ours in their classrooms and to adopt active learning approaches in large-scale computing education environments. 

\section*{Acknowledgments}

This material is based upon work supported by the National Science Foundation under Grant No. 2434428. Any opinions, findings, and conclusions or recommendations expressed in this material are those of the author(s) and do not necessarily reflect the views of the National Science Foundation.

The authors acknowledge the use of Grammarly and ChatGPT-4o, an AI-based language model developed by OpenAI, for spell-checking, grammar, and editing assistance. Both tools were used to refine the language and enhance the manuscript's clarity. No content, figures, or images were generated by AI for this work.

\bibliographystyle{IEEEtran}
\bibliography{references}

\end{document}